\documentclass[showpacs,preprint,preprintnumbers,amsmath,amssymb,floatfix,prd]{revtex4}

\usepackage{citesort}
\usepackage{epsfig}

\def\beq{\begin{equation}} 
\def\eeq{\end{equation}} 

\def\bea{\begin{eqnarray}} 
\def\eea{\end{eqnarray}} 
\def\bec{\begin{center}} 
\def\eec{\end{center}} 
\def\non{\nonumber}
\def\mc{\mathcal}
\def\mr{\mathrm}

\def \j{\mr{jet}} 

\newcommand{\vs}{\vspace*}

\begin{document}

\title{
Photoproduction of single inclusive jets at future $ep$~colliders 
in next-to-leading order QCD}
\author{B.\ J\"ager}
\affiliation{KEK Theory Division, Tsukuba 305-0801, Japan}

\begin{abstract}
A next-to-leading order QCD calculation for single-inclusive jet
photoproduction in unpolarized and longitudinally polarized lepton-hadron 
collisions is presented which consistently includes 
``direct'' and ``resolved'' photon contributions.  
The computation is performed within the ``small-cone approximation'' 
in a largely analytical form.
Phenomenological aspects of jet production at future $ep$~colliders 
such as the CERN-LHeC and the polarized BNL-eRHIC are discussed, 
placing particular emphasis on the perturbative stability of the 
predictions and the possibility to constrain the parton content of the photon. 
\end{abstract}


\maketitle

\section{Introduction}
%
Recent years have seen encouraging progress in the understanding of the inner
structure of hadrons, triggered by precision measurements at high energy
colliders, in particular the DESY-HERA. 
The knowledge gained on the dynamics which governs the interaction of 
color-charged particles is crucial  as new experiments are probing even higher
energies.  
The ultimate goal of the upcoming CERN-LHC, for instance, 
is the verification of the 
standard model by the discovery of a light Higgs boson or, else, 
the identification of new physics not anticipated within 
this framework. 
Subsequent measurements at a future linear collider will
constrain the parameters of the scenario realized in nature 
with even higher accuracy. 
However, poorly understood QCD effects may 
render the interpretation of new signatures in
terms of physics beyond the standard model difficult. 
The precise determination of the parton densities entering the description of 
any hadronic reaction is 
thus more timely than ever. 

Indeed, many improvements have been made recently on the parameterization 
of the proton distribution functions \cite{cteq6.6,mrsw,dssv}. 
Much less is known about the hadronic structure of the photon. 
This ignorance severely limits the predictive power of photon-induced reactions 
since the {\em resolved} contributions, which are associated with its hadronic
constituents, may be large at collider energies. 
As particularly promising
means to cure this deficiency, jet-production processes in the
photoproduction regime of lepton-hadron colliders, 
where the lepton beam acts as
source of quasi-real photons, have been identified 
due to large production rates and small systematic uncertainties.   
The capability of HERA to constrain the partonic structure of the photon 
as well
as the proton had been explored in great detail~\cite{klein} 
and several NLO-QCD calculations
for jet-photoproduction processes have become available,   
which turned out to
describe data reasonably well \cite{nlojets,owens}. For a polarization
upgrade of HERA, which has been discussed for some time \cite{pol-hera},  
a thorough sensitivity 
study has been performed in Ref.~\cite{svhera}. Polarized 
predictions at NLO-QCD accuracy have been
presented in \cite{deflo} in the form of a flexible Monte-Carlo program. 
 
Should single inclusive jets be used in the context of a ``global
analysis'' including photoproduction data, however, 
fast codes are essential which are based on analytical methods. 
The aim of this work is thus to present a calculation of single-inclusive jet
photoproduction in the framework of 
the ``small-cone approximation''~\cite{sca},
which allows for an entirely analytical computation of partonic hard-scattering
cross sections and has been demonstrated to approximate full jet cross sections
extremely well in related $pp$-scattering reactions \cite{sca:valid,jsv:jet}. 

Predictions can then be made for photoproduction cross sections at future
lepton-proton colliders such as the planned BNL-eRHIC \cite{erhic} 
and the CERN-LHeC \cite{lhec} which is currently under scrutiny. 
In principle, the code developed also allows for the
computation of jet-production observables at HERA. 
We refrain from presenting results for the HERA
kinematics here, however, since a vast number of NLO-QCD 
studies for this setting is
available in the literature \cite{nlojets,owens,deflo}.

The plan of the article is as follows: In Sec.~\ref{sec:technical} the technical
framework used will be specified. Section~\ref{sec:results} contains numerical
results for single-inclusive jet production at the LHeC and the eRHIC,
respectively. Conclusions will be given in Sec.~\ref{sec:conclusions}.

\section{Technical Framework}
\label{sec:technical}
%
We consider single-inclusive jet photoproduction in 
unpolarized and in longitudinally polarized 
lepton-proton collisions at NLO-QCD accuracy, i.e., the reaction 
$\ell p \to \ell' \mr{jet} X$. Provided the transverse momentum $p_T$ 
of the jet is large,
a polarized differential single-inclusive jet cross section can be written 
as a convolution
\bea
\label{eq:xsecdef}
d\Delta\sigma^{\ell p} &\equiv&
\frac{1}{2}\left[ d\sigma_{++} - d\sigma_{+-} \right] \non\\ 
&=& \sum_{a,b}
        \int dx_a \, dx_b 
        \,\Delta f^{\ell}_a(x_a,\mu_f)\,\Delta f^p_b(x_b,\mu_f)
        \non\\
&\times&
        d\Delta\hat{\sigma}_{ab\to \mr{jet}X}
        (S,x_a,x_b,\mu_r,\mu_f)
        \;,
\eea   
where the subscripts ``$++$'' and ``$+-$'' refer to the helicities of the
colliding leptons and protons, and $S$ is the available c.m.s.~energy squared. 
In Eq.~(\ref{eq:xsecdef}), $x_b$ denotes the momentum fraction of the proton
which is taken by parton $b$ and the $\Delta f^p_b(x_b,\mu_f)$ are the
longitudinally polarized parton distributions of the proton evaluated at a scale
$\mu_f$. 
The summation in Eq.~(\ref{eq:xsecdef}) is performed over all 
partonic channels $a+b\to \mr{jet}+X$ contributing to the reaction 
$\ell p \to \ell' \mr{jet} X$ with the corresponding spin-dependent cross
sections $d\Delta\hat{\sigma}_{ab\to \mr{jet}X}$, which are computed at NLO-QCD
accuracy. 

The photoproduction cross section $d\Delta\sigma$ 
which is observed in experiment consists of two
pieces: First, the {\em direct} part $d\Delta\sigma_\mr{dir}$, where a  
quasi-real photon emitted from the
lepton beam scatters off parton $b$ as an elementary particle such that
$a=\gamma$ in Eq.~(\ref{eq:xsecdef}).
Second, the {\em resolved} contribution $d\Delta\sigma_\mr{res}$, where 
the photon resolves into ``hadronic'' constituents, which in turn
interact with the partons emerging from the proton. 
In $d\Delta\sigma_\mr{res}$, $a$ denotes the parton stemming from the
photon.   
Direct and resolved contributions can be cast into the form of 
Eq.~(\ref{eq:xsecdef}) by defining $\Delta f^{\ell}_a$ as 
\beq
\label{eq:convol}
\Delta f_a^{\ell}(x_a,\mu_f)=
\int_{x_a}^1 \frac{dy}{y}\,\Delta P_{\gamma \ell}(y)
         \,\Delta f_a^\gamma\left(x_\gamma=\frac{x_a}{y},\mu_f\right),
\eeq
with 
\bea
\label{eq:weiz-will}
\nonumber
\Delta P_{\gamma \ell}(y)&=&\frac{\alpha_e}{2\pi}\Bigg\{
                \left[\frac{1-(1-y)^2}{y}\right]\ln\frac{
                Q_{\max}^2(1-y)}{m_{\ell}^2 y^2}\\
&+& 2m_{\ell}^2 y^2\left(\frac{1}{Q_{\max}^2}
                        -\frac{1-y}{m_{\ell}^2 y^2}\right)\Bigg\}
\eea
denoting the spin-dependent 
Weizs\"acker-Williams ``equivalent-photon'' spectrum
for the emission of a circularly polarized collinear photon with a 
virtuality smaller than 
$Q_{\max}^2$ by a lepton of mass $m_{\ell}$ \cite{deflo}.
For the direct contribution, $x_a$ has to be identified with the momentum
fraction $y$ of the lepton which is taken by the photon and thus
\beq
\Delta f_a^\gamma = \delta(1-x_\gamma)\,.
\eeq
In the resolved case, the $\Delta f_a^\gamma$ denote the parton distributions of
the circularly polarized photon which are completely unmeasured so far. 

It is important to note that beyond the leading order 
neither $d\Delta\sigma_\mr{dir}$ nor
$d\Delta\sigma_\mr{res}$ are measurable cross sections {\em per~se}, as their
individual values depend on the factorization scheme chosen. Only if both pieces
are evaluated using the same factorization prescription, their sum 
\beq
d\Delta\sigma = d\Delta\sigma_\mr{dir}+d\Delta\sigma_\mr{res}\,
\eeq
is a meaningful quantity.  
The unpolarized jet cross section, $d\sigma=[d\sigma_{++} + d\sigma_{+-}]/2$, 
is obtained in complete analogy to the 
polarized one by replacing all spin-dependent parton distributions and partonic
cross sections with their spin-averaged counterparts. The spin-averaged
equivalent-photon spectrum can be found in Ref.~\cite{unp-ww}.
  
In order to compute the  hard-scattering cross sections
$d(\Delta)\hat{\sigma}_{ab\to \mr{jet}X}$, an algorithm has to be specified
describing the formation of jets by the 
partons which undergo the hard scattering. 
A frequently adopted choice is to
define a jet as the deposition of the total transverse energy  
of all final-state partons that fulfill
\beq
(\eta-\eta^i)^2+(\phi-\phi^i)^2\leq R^2\,,
\eeq
where $\eta^i$ and $\phi^i$ denote the pseudo-rapidities and azimuthal angles of
the  particles and $R$ the jet cone aperture. The jet variables are defined as
\bea
E_T &=& \sum_i E_T^i\,, \non\\
\eta &=& \sum_i \frac{E_T^i\eta^i}{E_T} \,,\\
\phi &=& \sum_i \frac{E_T^i\phi^i}{E_T} \,.\non
\eea

We resort to the so-called ``small-cone approximation'' \cite{sca},  
which can be considered as an expansion of the jet cross
section in terms of 
$R$ 
of the form $\mc{A}\log{R} + \mc{B} + \mc{O}(R^2)$. 
Neglecting $\mc{O}(R^2)$ pieces, the evaluation and phase-space integration of
the partonic 
cross sections can be performed analytically, as is our intention. 
The small-cone approximation has been shown~\cite{sca:valid,jsv:jet} to account extremely well for 
jet observables up to cone sizes of 
about $R\approx 0.7$ by explicit comparison to
calculations that take $R$ fully into account.  
The predictions of Ref.~\cite{jsv:jet} for 
single-inclusive jet production at the BNL-RHIC within
the small-cone approximation are in excellent agreement with recent data from the STAR
collaboration \cite{star:jets}.

Since the resolved contribution $d(\Delta)\sigma_\mr{res}$ 
is technically equivalent to the 
jet production cross section in hadronic collisions, $pp\to \j X$, 
the corresponding partonic  matrix elements squared can be taken from this 
previous calculation \cite{jsv:jet}.
The direct contributions are adapted 
from the results for the $d(\Delta)\hat{\sigma}_{\gamma b\to cX}$  in  
single-inclusive hadron photoproduction \cite{jsv:erhic,jsv:compass} 
with the techniques of Ref.~\cite{jsv:jet}. 

The hard-scattering cross sections are then implemented in a Monte-Carlo program which
performs the convolutions with the parton distributions of the proton and with
the equivalent photon spectrum numerically by means of an adaptive {\tt VEGAS}
integration. 

As consistency check, we have calculated jet cross sections for HERA
kinematics and compared our results successfully 
to those of Refs.~\cite{svhera,deflo,owens}.

\section{Numerical Results}
\label{sec:results}
%
We now turn to a phenomenological study of jet production at future 
lepton-hadron colliders. Our aim is twofold: First, we will discuss the impact
of NLO corrections on transverse momentum and rapidity distributions in various
kinematic regimes and explore the stability of our predictions with respect to
scale variations.  Second, we will investigate the sensitivity of jet-production
cross sections to the parton distribution functions of the photon and the proton
and demonstrate how to increase the impact of specific contributions by
adjusting parameters of the analysis. 

\subsection{Single-inclusive jet production at the LHeC}
\label{sec:lhec}
%
New opportunities for the measurement of single-inclusive jets in lepton-hadron
collisions could be provided by a future LHeC $ep$~collider at CERN, which is
currently under scrutiny~\cite{lhec}. 
In the following we will present predictions  for the 
scenario where an electron
beam circulates in the existing LHC tunnel with a nominal energy of
$E_e=70$~GeV, which in conjunction with the 7~TeV proton beam gives 
rise to $ep$ collisions with a c.m.s.~energy of
$\sqrt{S}=1.4$~TeV. 

For the equivalent-photon approximation \cite{unp-ww} 
similar parameters as for the H1 and ZEUS experiments
at HERA are used, $Q_\mr{max}^2=1$~GeV$^2$ and $0.2\leq y\leq 0.85$.  
All LO (NLO) calculations are performed with the CTEQ6L (CTEQ6M) parton
distributions \cite{cteq6} and the according one-loop (two-loop) values for the strong coupling constant
$\alpha_s$ in the $\overline{\mr{MS}}$-factorization scheme. 
For the parton distribution functions of the photon the 
LO (NLO)  GRV set~\cite{grv} is employed. 

\begin{figure*}[t!]
\bec
\epsfig{figure=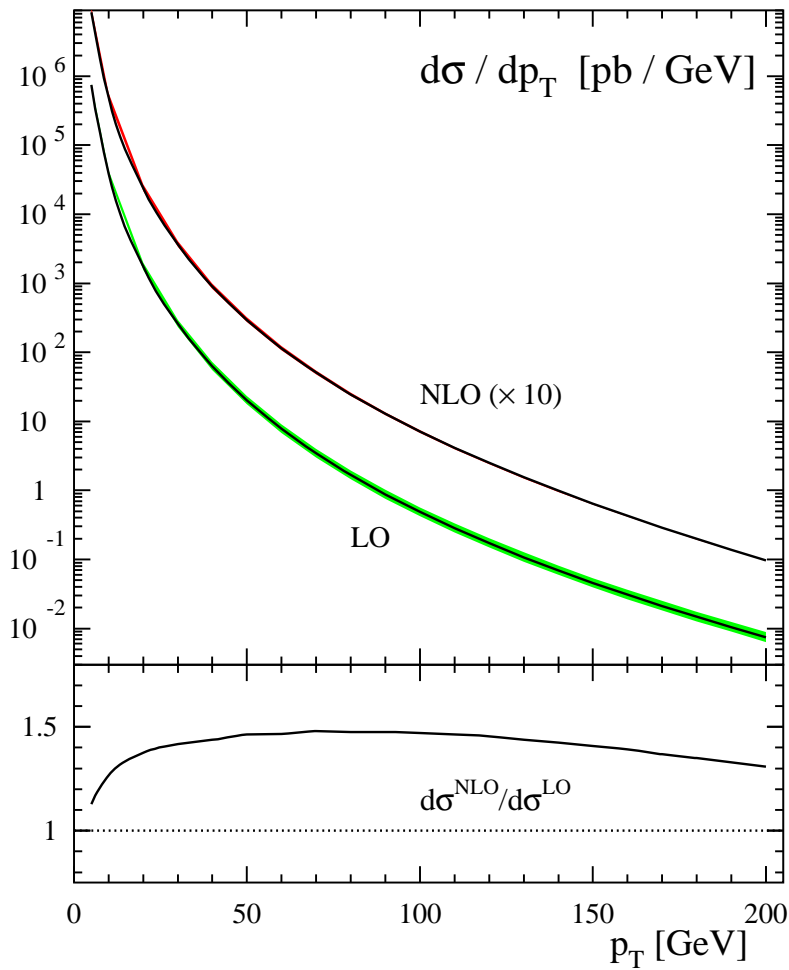,width=0.75\textwidth}
\eec
\vs{-0.7cm}
\caption{
\label{fig:lhec-pt-scdep}
Single-inclusive jet cross section at the LHeC as function of $p_T$ 
in NLO (multiplied by a factor of 10) and LO, integrated over $-1<\eta<4$. 
The shaded bands correspond to a scale variation of the NLO and LO
results, respectively, in the
range $p_T/2\leq \mu_r=\mu_f\leq 2 p_T$. The lower panel shows the associated 
K~factor. }
\end{figure*}

The major motivation for studying jet cross sections beyond leading order 
is to reduce the theoretical uncertainties associated with a tree-level calculation.
To access the improvement which can be gained by the inclusion of NLO corrections
the single-inclusive jet cross section at the
LHeC is shown in Fig.~\ref{fig:lhec-pt-scdep} 
%
as function of the jet transverse momentum $p_T$ at LO and NLO.
The jet cone size is set to $R=0.7$ and rapidities are integrated over
$-1<\eta<4$ in the laboratory frame. The solid lines correspond to the setting $\mu_r=\mu_f=p_T$. 
The bands have been obtained by varying the factorization and renormalization
scales simultaneously in the range  $p_T/2\leq \mu_r=\mu_f\leq 2 p_T$.
Throughout the $p_T$ interval considered, 
the scale dependence is very small at LO
already, amounting to about $20\div 30$~\%. 
While at
low values of $p_T$ the NLO corrections do not significantly improve the scale
dependence of the cross section, towards higher $p_T$ the scale uncertainty of
the NLO result is extremely small, going down to the level of 2.5~\% at
$p_T=200$~GeV. 
This behavior indicates that the perturbative expansion is
under excellent control, provided single-inclusive jets are produced at high transverse momentum.  
The impact of the NLO corrections on the cross section is indicated by the
K~factor, which we define as
\beq
\label{eq:kfac}
K(x) = \frac{d\sigma^\mr{NLO}/dx}{d\sigma^\mr{LO}/dx}\,.
\eeq
For $\mu_r=\mu_f=p_T$, $K(p_T)$ is larger than one everywhere, 
which reflects the increase of
the cross section by the inclusion of the NLO contributions. 

\begin{figure*}[tp]
\bec
\epsfig{figure=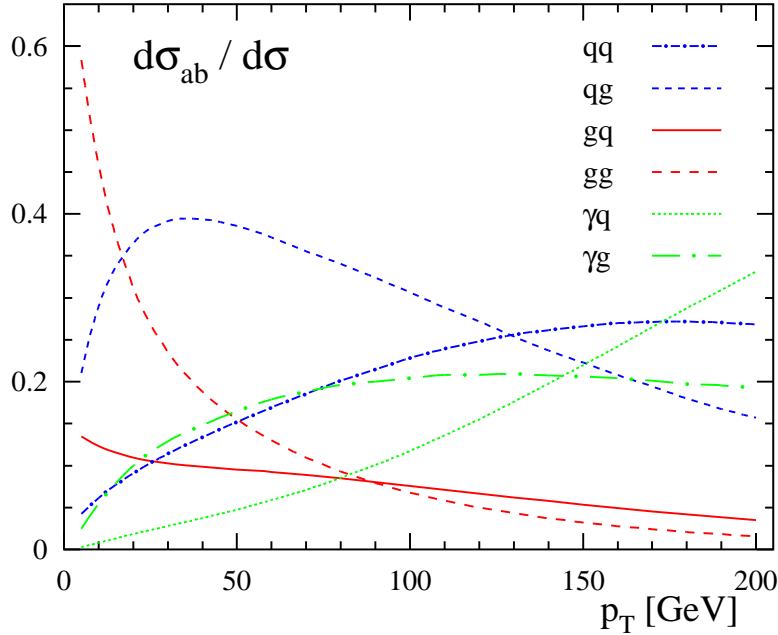,width=0.75\textwidth,clip=}
\eec
\vs{-0.7cm}
\caption{
\label{fig:lhec-pt-sub}
Relative contributions of different partonic subprocesses $ab \to \j X$ to
the NLO single-inclusive jet cross section at the LHeC, 
integrated over $-1<\eta<4$.  }
\end{figure*}

Figure~\ref{fig:lhec-pt-sub} 
%
illustrates the contributions of different partonic subprocesses to the
rapidity-integrated NLO cross section. The
resolved contributions are dominant over a large range of $p_T$ with the 
direct contributions starting to take over only at $p_T\approx 200$~GeV. 
Since the cross section is largest at low values of
$p_T$, this indicates that single-inclusive jet production at LHeC energies 
offers excellent
opportunities for a more accurate determination of the parton content of the
photon.

\begin{figure*}[tp]
\bec
\epsfig{figure=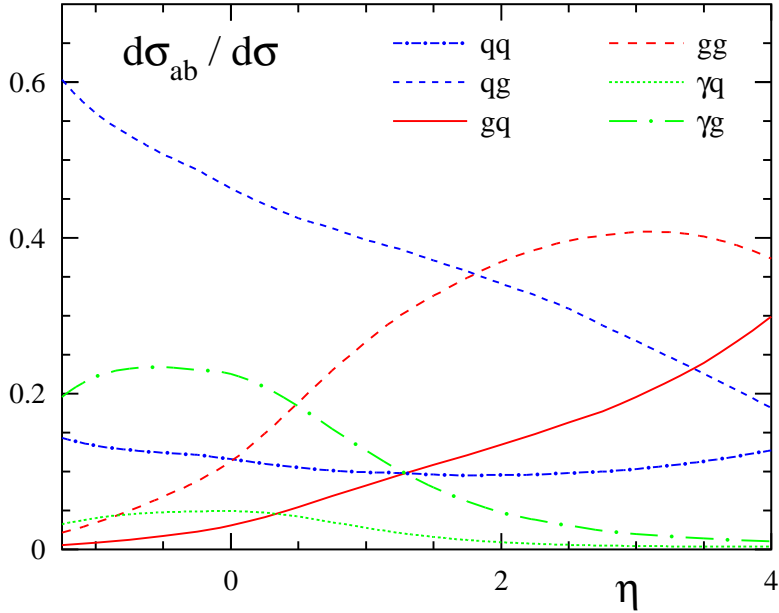,width=0.75\textwidth,clip=}
\eec
\vs{-0.7cm}
\caption{
\label{fig:lhec-eta-sub}
Relative contributions of different partonic subprocesses $ab \to \j X$ to
the NLO single-inclusive jet cross section at the LHeC, 
integrated over $p_T>20$~GeV.  }
\end{figure*}

To this end, the study of rapidity-differential cross sections is
particularly suitable, since the momentum fractions of the hadronic constituents
of the photon and the proton
can be considered as functions of the rapidity of the observed jet 
in the laboratory frame, $\eta$.  
As explained, e.g., in
Ref.~\cite{svhera}, if counting positive rapidity in the forward direction 
of the proton, large $x_\gamma\to 1$ are probed at large negative values 
of $\eta$. In 
this region, the direct contribution is expected to be largest 
and the $f^\gamma$ are dominated by the purely
perturbative ``pointlike'' part which does not depend on the hadronic structure
of the photon.  
The relative contributions of the various direct and resolved channels to the
full cross section are depicted as function of rapidity  
in Fig.~\ref{fig:lhec-eta-sub}. 
%

\begin{figure*}[t]
\bec
\epsfig{figure=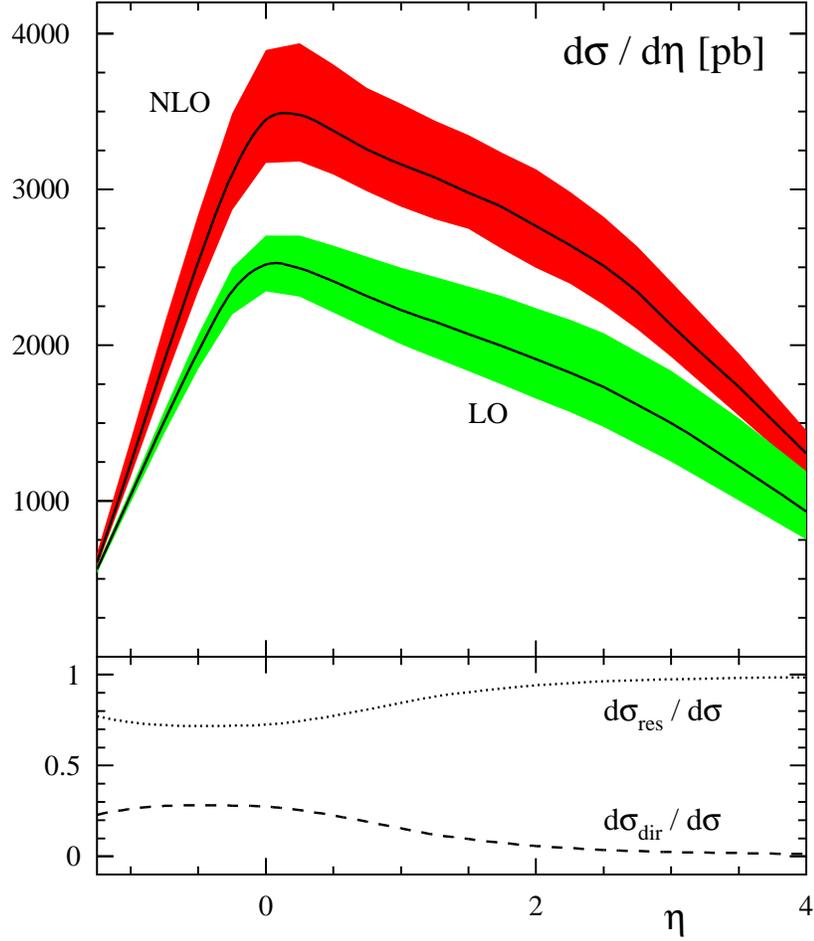,width=0.75\textwidth,clip=}
\eec
\vs{-0.7cm}
\caption{
\label{fig:lhec-eta-scdep}
Single-inclusive jet cross section at the LHeC as function of $\eta$ 
in NLO and LO, integrated over $p_T>20$~GeV. 
The shaded bands correspond to a scale variation in the
range $p_T/2\leq \mu_r=\mu_f\leq 2 p_T$. 
The lower panel depicts the relative contributions of the 
direct and resolved sub-processes to the NLO cross section. 
}
\end{figure*}

\begin{figure*}[t]
\bec
\epsfig{figure=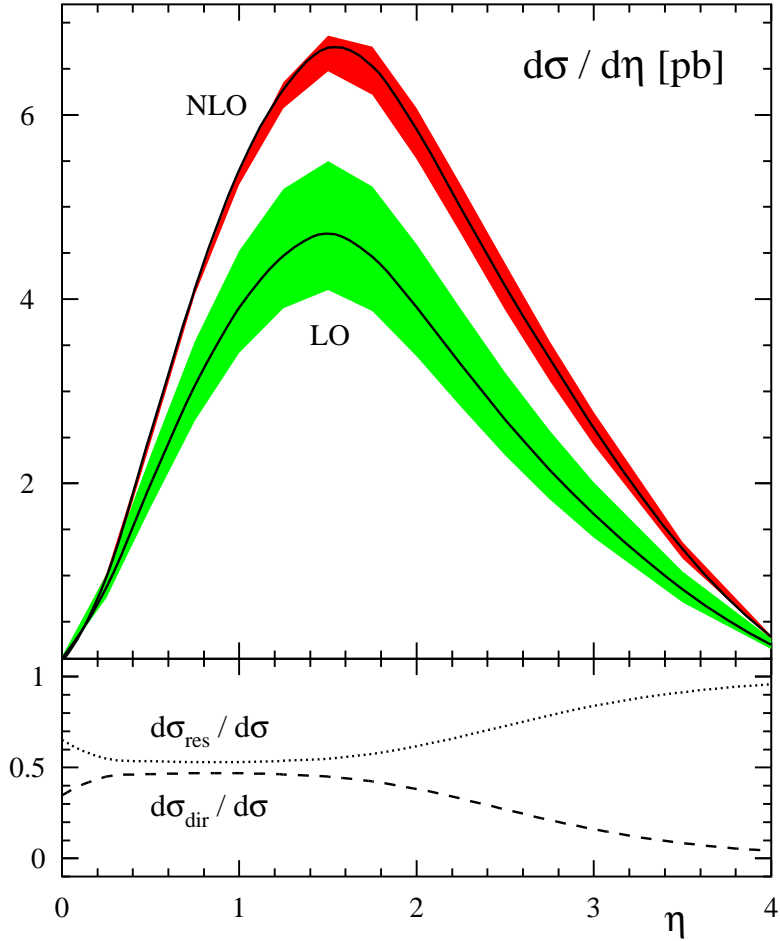,width=0.75\textwidth,clip=}
\eec
\vs{-0.7cm}
\caption{
\label{fig:lhec-eta-scdep-cut} 
Same as in Fig.~\ref{fig:lhec-eta-scdep}, but now integrated over 
$p_T>100$~GeV.  }
\end{figure*}

The LO and NLO rapidity-dependent jet cross
sections integrated over $p_T>20$~GeV are shown  in
Fig.~\ref{fig:lhec-eta-scdep} 
%
along with the
associated scale uncertainties, obtained by varying renormalization and
factorization scales simultaneously in the range $p_T/2\leq \mu_r=\mu_f\leq 2
p_T$. 
Also shown are the relative contributions of the direct and the resolved cross
sections at NLO. 
Strikingly, the 
scale uncertainty of the rapidity distribution is not improved by the 
inclusion of NLO corrections. This feature can be traced back to the large
weight of contributions from relatively low $p_T$, where the LO and the NLO 
scale dependences are of similar size, c.f.~Fig.~\ref{fig:lhec-pt-scdep}. 
After imposing a transverse momentum cut of $p_T>40$~GeV, the 
scale dependence is slightly
improved with the size of the cross section being reduced at the same time 
by approximately one order of magnitude.  Even smaller scale uncertainties 
are obtained for $p_T>100$~GeV, see  
Fig.~\ref{fig:lhec-eta-scdep-cut}. 
%
As the transverse momentum cut is increased, the 
impact of the direct photon contributions becomes larger and 
the resolved sub-processes start to be less important.

\subsection{Polarized photoproduction of single-inclusive jets at eRHIC}
%
Photoproduction experiments complimentary to the measurements possible at the
LHeC could be performed at the future lepton-proton collider eRHIC at BNL
\cite{erhic}. 
The polarized beams available at eRHIC 
offer unique opportunities for studying the spin structure of the circularly
polarized photon, which is completely unknown so far. 

In the following we assume a hadronic c.m.s.~energy of  $\sqrt{S}=100$~GeV. For
the equivalent-photon approximation \cite{unp-ww,deflo} we choose
$Q_\mr{max}^2=1$~GeV$^2$ and $0.2\leq y\leq 0.85$. 
In the unpolarized case, we stick to the 
parton distributions of Sec.~\ref{sec:lhec}. 
For the spin-dependent proton
distribution functions we use the GRSV standard set \cite{grsv2000} as default, since it
contains both, a LO and an NLO parameterization. 
To illustrate the impact of different parton densities, we 
will also employ the new DSSV set \cite{dssv}, 
which is only available at NLO, however. 
The parton distributions of the polarized photon are completely unmeasured so
far. We therefore consider the two extreme scenarios of Ref.~\cite{sv} 
with minimal [$\Delta f^\gamma(x,\mu_0)=0$] and maximal 
[$\Delta f^\gamma(x,\mu_0)= f^\gamma(x,\mu_0)$] 
saturation of the positivity constraint 
$|\Delta f^\gamma(x,\mu_0)|\leq f^\gamma(x,\mu_0)$ \cite{lomodels}. 
Here, $\mu_0$ denotes the
scale where the boundary conditions for the evolution are fixed.
If not specified otherwise, the ``maximal'' set will be used.   
 
\begin{figure*}[t!]
\bec
\epsfig{figure=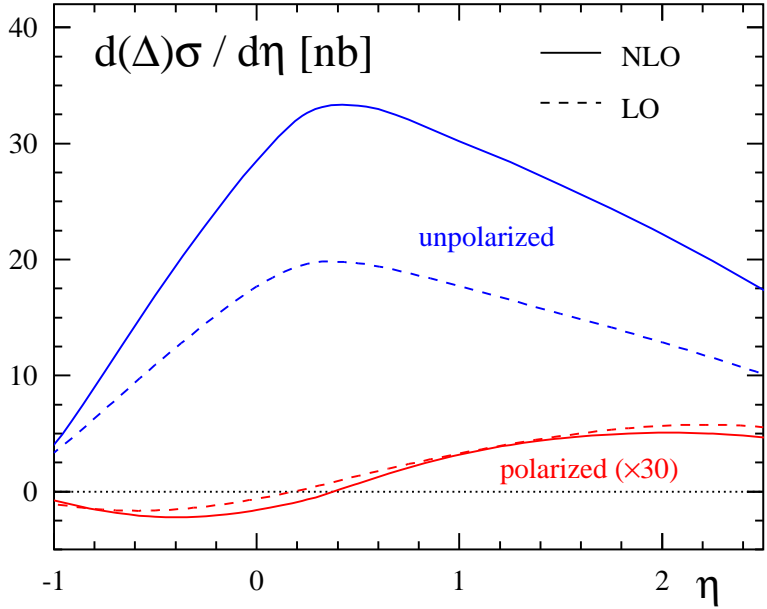,width=0.75\textwidth,clip=}
\eec
\vs{-0.7cm}
\caption{
\label{fig:erhic-xsec}
Polarized and unpolarized single-inclusive jet cross sections at eRHIC 
as function of $\eta$ 
in NLO  and LO, integrated over $p_T>4$~GeV. The polarized cross section is
multiplied by a factor of~30. }
\end{figure*}

Figure~\ref{fig:erhic-xsec}
%
shows the rapidity dependent unpolarized and polarized single-inclusive jet 
cross sections at LO and NLO integrated over $p_T>4$~GeV. 
The spin-averaged cross section receives sizeable positive
NLO corrections resulting in $1.2\lesssim K(\eta) \lesssim 1.7$. In
the spin-dependent case, the NLO effects are rather small 
yielding a K~factor close to one over a large range in $\eta$. 

\begin{figure*}[t!]
\bec
\epsfig{figure=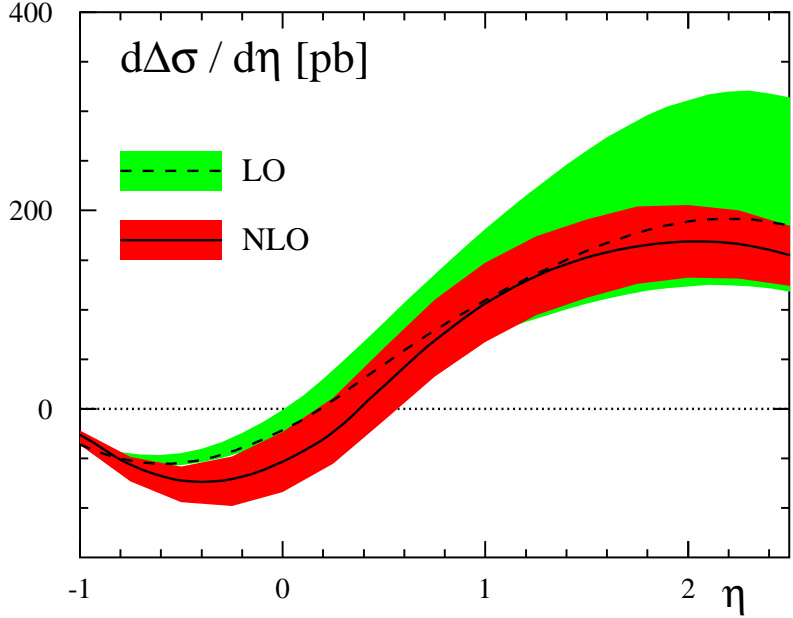,width=0.75\textwidth,clip=}
\eec
\vs{-0.7cm}
\caption{
\label{fig:erhic-eta-scdep}
Single- inclusive jet cross section at eRHIC as function of $\eta$ 
in NLO and LO, integrated over $p_T>4$~GeV. 
The shaded bands correspond to a scale variation in the
range $p_T/2\leq \mu_r=\mu_f\leq 2 p_T$.   
}
\end{figure*}

The scale uncertainty of the polarized cross section, illustrated by
Fig.~\ref{fig:erhic-eta-scdep}, 
%
is sizably reduced at NLO. Scale dependences are 
generally smaller for jet cross sections than for 
related hadron-production observables, such as the 
pion-production cross sections at eRHIC discussed in  Ref.~\cite{jsv:erhic}. 
This hierarchy is also observed in related $pp$-scattering reactions (see
Refs.~\cite{jssv:pions,jsv:jet}) and 
is mainly due to the presence of scale dependent fragmentation
functions in processes with identified hadrons. 
These non-perturbative objects describe 
the formation of hadrons from the final-state partons
in a hard-scattering process at a specific scale, which may be
different from the scale at which initial-state singularities are factorized
into the distribution functions of the photon and proton, respectively.

\begin{figure*}[t!]
\bec
\epsfig{figure=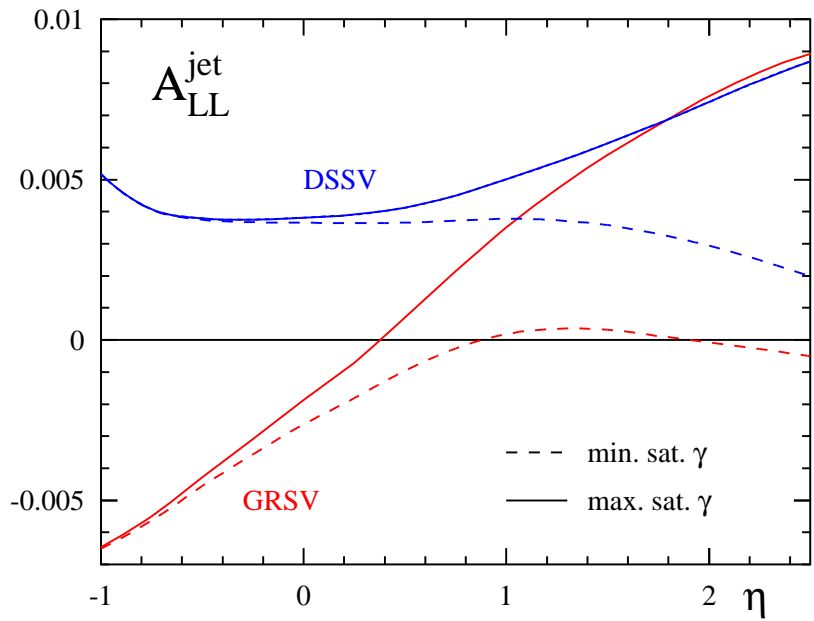,width=0.75\textwidth,clip=}
\eec
\vs{-0.7cm}
\caption{
\label{fig:erhic-asy}
NLO-QCD spin asymmetry for single-inclusive jet photoproduction 
at eRHIC integrated over $p_T>4$~GeV for two different choices 
of proton distribution functions and
the two extreme sets of polarized photon densities. 
}
\end{figure*}

For extracting information on the hadronic structure of the circularly
polarized photon the experimentally accessible spin asymmetry 
\beq
A_\mr{LL}^\j = \frac{d\Delta\sigma}{d\sigma} 
             = \frac{d\sigma_{++}-d\sigma_{+-}}{d\sigma_{++}+d\sigma_{+-}}
\eeq
is most suitable.  	 
At large positive rapidities, $A_\mr{LL}^\j$ is particularly sensitive to the
parton content of the resolved photon as exemplified 
in Fig.~\ref{fig:erhic-asy},     
%
where the spin asymmetry is shown for the two extreme sets of polarized photon
densities introduced above. 
To demonstrate that this sensitivity is not obscured by our currently
incomplete knowledge of the polarized proton, predictions
are made for two different sets of proton distributions which mainly differ in
the parameterization of the polarized gluon density. 
Towards negative values of $\eta$,  $A_\mr{LL}^\j$ becomes 
less sensitive to the
hadronic structure of the photon. With the $qg$ subprocess being
dominant in this region, the spin asymmetry could
provide new information on the gluon density of the proton.

\section{Conclusions}
\label{sec:conclusions}
%
In this work an NLO-QCD calculation of single-inclusive jet photoproduction in
unpolarized and longitudinally polarized lepton-hadron 
collisions has been presented. The
computation was performed in the context of the small-cone approximation. 
In this way, the evaluation of the 
partonic matrix elements and large parts of the phase-space integration 
could be performed analytically, yielding a stable and much faster computer code
than comparable programs that are based on a purely numerical approach. 

We have performed a phenomenological analysis of jet photoproduction at future
$ep$~colliders, in particular the LHeC and eRHIC. 
We found that NLO corrections are sizeable at the LHeC with
K~factors of about 1.5. In the low-to-moderate $p_T$ regime the resolved photon
contributions are by far dominant and scale uncertainties are large even
beyond the leading order. If a large transverse momentum cut is  imposed, 
these uncertainties can
be brought down. At high $p_T$, quark-initiated processes dominate over the
gluonic channels. 

For eRHIC, we have focused on jet 
photoproduction by longitudinally polarized beams.
Polarized cross sections exhibit smaller NLO corrections than their spin-averaged
counterparts with K~factors close to one and only moderate scale dependences.
The experimentally relevant double-spin asymmetry exhibits
pronounced sensitivity to the hadronic structure of the (resolved) photon in the
positive-rapidity regime. At negative values of $\eta$,  $A_\mr{LL}^\j$ could
yield further information on the gluon polarization of the proton complementary
to experiments in hadron-hadron collisions. 

In summary, future lepton-hadron colliders offer new possibilities for further 
constraining the hadronic structure of the (real) photon in the unpolarized
case, and unique opportunities to access the completely unknown parton
distributions of the polarized photon. Embedded in a global analysis of hadronic
scattering reactions, single-inclusive jet production 
will also help to further pin
down the parton densities of the proton, in particular its gluonic component.

\acknowledgments
I am grateful to  Marco Stratmann for valuable discussions and comments.
This  work was supported by the Japan Society for the Promotion of
Science (JSPS).


\end{document}